\documentclass{ifacconf}

\usepackage{cite}
\usepackage{amsmath,amssymb,amsfonts}
\usepackage{subfigure}
\usepackage{graphicx}
\usepackage{textcomp}
\usepackage{xcolor}      
\usepackage{natbib}
\usepackage{algpseudocode}
\usepackage{algorithmicx,algorithm}
\usepackage{pgf}
\usepackage{tikz}
\usetikzlibrary{positioning, arrows.meta,automata}
\usetikzlibrary{calc}
\usepackage[latin1]{inputenc}

\newcommand{\EQ}{\begin{eqnarray}}
\newcommand{\EN}{\end{eqnarray}}
\newcommand{\EQQ}{\begin{eqnarray*}}
\newcommand{\ENN}{\end{eqnarray*}}
\newcommand{\EF}{\begin{flalign}}
\newcommand{\EFN}{\end{flalign}}
\newcommand{\BM}{\left[\begin{array}}
\newcommand{\EM}{\end{array}\right]}

\newcommand{\bdefinition}{\begin{definition} \begin{rm} }
\newcommand{\edefinition}{ \end{rm} 
\end{definition} }
\newcommand{\bremark}{\begin{remark} \begin{rm} }
\newcommand{\eremark}{ \end{rm} 
\end{remark} }
\newcommand{\btheorem}{\begin{theorem}  \begin{rm} }
\newcommand{\etheorem}{ \end{rm} 
\end{theorem} }
\newcommand{\blemma}{\begin{lemma} \begin{rm} }
\newcommand{\elemma}{ \end{rm} 
\end{lemma} }
\newcommand{\bcorollary}{\begin{corollary} \begin{rm} }
\newcommand{\ecorollary}{ \end{rm} 
\end{corollary} }
\newcommand{\bproposition}{\begin{proposition} \begin{rm} }
\newcommand{\eproposition}{ \end{rm} 
\end{proposition} }
\newcommand{\bclaim}{\begin{claim} \begin{rm} }
\newcommand{\eclaim}{ \end{rm} 
\end{claim} }
\newcommand{\bproblem}{\begin{problem} \begin{rm} }
\newcommand{\eproblem}{ \end{rm} 
\end{problem} }
\newcommand{\bexam}{\begin{exam}  }
\newcommand{\eexam}{\end{exam} }

\newtheorem{theorem}{Theorem}
\newtheorem{lemma}{Lemma}
\newtheorem{exam}{Example}
\newtheorem{corollary}{Corollary}
\newtheorem{definition}{Definition}
\newtheorem{proposition}{Proposition}
\newtheorem{problem}{Problem}
\newtheorem{remark}{Remark}[section]

\begin{document}

\def \L{\mathcal{L}}
\def \OO{\Sigma_{o}}
\def \UO{\Sigma_{uo}}
\def \OI{\Sigma_{a}}
\def \UI{\Sigma_{ua}}
\def \IPO{P_o^{-1}}
\def \IPI{P_a^{-1}}
\def \XO{X_{\obs}}
\def \deltao{\delta_{\obs}}
\def \xoi{x_{\obs,0}}
\def \Obs{\text{Obs}}
\def \obs{{o}}
\def \spec{\text{spec}}
\def \XI{X_{\int}}
\def \deltai{\delta_{\int}}
\def \xii{x_{\int,0}}
\def \Int{\text{Int}}
\def \int{{a}}

\def \xivi{x_{\text{int},0}}

\def\XY#1{{\textcolor{red}{ {\bf XY:} #1}}}

\begin{frontmatter}

\title{You Don't Know What I Know: \\ 
On Notion of High-Order Opacity in Discrete-Event Systems\thanksref{footnoteinfo}} 

\thanks[footnoteinfo]{This work was supported by the National Natural Science Foundation of China (62061136004, 62173226, 61803259) and by Shanghai Jiao Tong University Scientific and Technological Innovation Funds.}

\author[First]{Bohan Cui} 
\author[First]{Xiang Yin}
\author[First]{Shaoyuan Li}
\author[Second]{Alessandro Giua}

\address[First]{Department of Automation, Shanghai Jiao Tong University, Shanghai, 200240, China\\ (E-mail: \{bohan\_cui, yinxiang, syli\}@sjtu.edu.cn).}
\address[Second]{DIEE, University of Cagliari, Cagliari 09123, Italy.\\ 
(E-mail: giua@unica.it.)}

\begin{abstract}                
In this paper, we investigate a class of information-flow security properties called \emph{opacity} in partial-observed discrete-event systems. Roughly speaking, a system is said to be opaque if the intruder, which is modeled by a passive observer, can   never   determine the ``secret" of the system for sure. Most of the existing notions of opacity consider secrets related to the actual behaviors of the system.  In this paper, we consider a new type of secret related to the \emph{knowledge of the system user}.  
Specifically, we assume that the system user also only has partial observation of the system and has to reason the actual behavior of the system.  We say a system is \emph{high-order opaque} if the intruder can never determine that the system user knows some information of importance based on its own incomparable information. We provide the formal definition of high-order opacity. Two algorithms are provided for the verification of this new notion: one with doubly-exponential complexity for the worst case and the other with single-exponential complexity. Illustrative examples are provided for the new notion of high-order opacity. \end{abstract}
\begin{keyword}
Opacity, Discrete Event Systems, Partial Observation.
\end{keyword}

\end{frontmatter}

\section{Introduction} 
With the development of information and network technologies, smart devices with both computation and communication capabilities have been widely used in  cyber-physical control systems.  The large information transmission between devices, on the one hand, makes control systems more flexible and intelligent by enabling, for example, edge or cloud based control architectures. On the other hand, however, the large amount of communications also makes security issue much more severe. Therefore, security and privacy considerations have been becoming increasingly more important in the analysis and design of networked cyber-physical systems (CPS).  

In this work, we consider an important class of information-flow security properties, called \emph{opacity}, in the context of discrete-event systems (DES), a widely used formal model for describing high-level behaviors of CPS \citep{cassandras2021introduction}. Generally speaking, opacity captures whether or not some secret behaviors of the system can be revealed to an external intruder that can access partial information-flow of the system. A system is said to be opaque if the intruder can never infer the secret behaviors of the system based on its own observation. 
It has been shown that the notions of opacity subsume many existing security properties in the literature  \citep{bryans2008opacity}. 
Due to its importance, verification and enforcement of opacity  has drawn considerable attention in the past few years; see, e.g., \cite{lafortune2018history,behinaein2019optimal,liu2021compositional,balun2021comparing,wintenberg2022general}.
 
Since opacity is an information-flow security property,  one of the key ingredients in its definition is what is the ``secret" the system wants to hide against the intruder. 
The most general formulation of opacity is language-based opacity, where secret is modeled as a set of secret strings; see, e.g., \cite{lin2011opacity}. 
In specific applications, however, secret strings usually have concrete meanings, e.g., visited  a critical location at some instant. Therefore, state-based secret have been more widely adopted in the verification and synthesis of opacity. State-based opacity includes, e.g., initial-state opacity \citep{saboori2013verification,lai2021initial}, current-state opacity \citep{tong2017verification}, infinite-step opacity \citep{saboori2011inf,yin2017new} and $K$-step opacity \citep{yin2019infinite,ma2021verification}. 

All of the aforementioned notions of opacity consider the secret of system as some \emph{actual behaviors of importance}, e.g., is visiting or has visited a secret state.  In some applications, however, the secret of the system may be \emph{the knowledge of the user} about the current status of the system. 
To motivate the ``knowledge security" issue, let us consider the following scenario. Suppose that there are two investors in the market. Each of them knows the trading strategy of the other but does not the information available to the other. Suppose that one investor wants to investigate some undisclosed information about a stock in order to decide to buy it or not. At the same time, he does not want the other investor to know the fact that he has obtained sufficient information for trading; otherwise, the other investor may take the advantage of the information that the first investor will buy the stock. This example suggests that, in some scenarios, knowing something of importance is also closely related to the security consideration. 

In this paper, we investigate the verification of opacity from a new perspective by considering the secret of the system as the knowledge of the system user. Specifically, we model the user and the intruder as two observers with different information-flow. 
The objective of the user is to gain sufficient knowledge about the system, e.g., for the purpose of decision-making. Here we consider a specific type of knowledge objective, which is to distinguish  certain pairs of states \citep{wang2007algorithm,sears2014computing}. 
On the other hand, the intruder may use its own information to infer whether or not the user has sufficient knowledge.  We formulate the knowledge-security of the user in terms of the notion of \emph{high-order opacity}. Specifically, the system user is said to be high-order opaque if the intruder can never determine for sure whenever the user knows something of importance. We present two algorithms for verifying high-order opacity. The first algorithm is based on the construction of two successive observers, which yields a doubly-exponential complexity. The second algorithm is based on estimating, from the intruder's point of view,  all state pairs that cannot be distinguished by the user; the complexity of the second approach is only single-exponential in the size of the plant. 

The formulated new notion of high-order opacity is conceptually related to the notion of HyperLTL \citep{finkbeiner2015algorithms} in the context of model checking and the notion of dynamic epistemic logic \citep{van2007dynamic} in the context of knowledge theory. In principle, both of them allow the descriptions such as one agent knows that the other agent does not know something. However, to our knowledge, the issue of knowledge security has not yet been considered explicitly. Here, we formulate this new security consideration directly using high-order opacity and develop effective customized verification algorithms in the context of DES.  Our work is also related to inference-based decision-making in the context of decentralized DES, where agents need to infer the knowledge of other agents; see, e.g., \cite{yoo2004decentralized,ricker2007knowledge,takai2010inference}. However, these works consider a collaborative setting, where agents need to complete a global task, but here we consider a security problem, where the objective of the intruder is antagonistic. 

The rest of this paper is organized as follows. Section \ref{sec-pre} provides some basic preliminaries. In Section \ref{sec-notion}, we first provides a motivating example, and  then propose the notion of \emph{high-order opacity}. Two effective verification algorithms for high-order opacity are presented in Section \ref{sec-verify}. Finally, we conclude the paper and discuss the future directions in Section \ref{sec-con}.

\section{Preliminaries}\label{sec-pre}

\subsection{System Model}

Let $\Sigma$ be a finite set of events. 
A string is a finite sequence of events and $\Sigma^*$ denotes the set of all strings over $\Sigma$ including the empty string $\epsilon$. For any string $s\in \Sigma^*$, $|s|$ denotes the length of $s$ with $|\epsilon|=0$. A language $L\subseteq\Sigma^*$ is a set of strings and we denote by $\bar{L}$ the prefix-closure of language $L$, i.e., $\bar{L}=\{s\in\Sigma^*:\exists w\in\Sigma^* \text{ s.t. }  sw\in L\}$.

We consider a DES modeled by a  deterministic finite-state automaton (DFA)\vspace{1pt}
\[
G=(X,\Sigma,\delta,x_0), \vspace{3pt}
\]
where $X$ is a finite set of states,  
$\Sigma$ is a finite set of events,  
$\delta:X \times \Sigma \to X$ is the partial transition function, where for any $x,x'\in X$, $\sigma\in\Sigma$, $x'=\delta(x,\sigma)$ means that there exists a transition from state $x$ to state $x'$ via event $\sigma$ and  $x_0\in X$ is the initial state. 
The transition function is also extended to $\delta: X\times \Sigma^* \to X$ recursively by: 
 (i)  for any $x\in X$, $\delta(x,\epsilon)=x$ and 
 (ii) for any $x\in X, s\in \Sigma^*, \sigma \in \Sigma$, we have $\delta(x,s\sigma)= \delta( \delta(x,s) ,\sigma)$. 
The set of all strings  generated by $G$ starting from state $x\in X$ is defined as $\L(G,x)=\{s \in \Sigma^*:\delta(x,s)!\}$, where ``$!$" means ``is defined". The set of all strings generated by $G$ is defined as $\L(G):=\L(G,x_0)$. For any $s\in\L(G)$,  we write $\delta(x_0,s)$ simply as $\delta(s)$. For the sake of simplicity, we assume that system $G$ is live, i.e., for any $x\in X$, there exists $\sigma\in \Sigma$ such that $\delta(x,\sigma)!$. 
 
\begin{remark}
Here we consider DES with deterministic transitions with a unique initial state. This is without loss of generality in the partial-observation setting because it is well-known that one can always use unobservable events to mimic non-determinism. 
\end{remark}

Let $\Sigma'\subseteq \Sigma$ be a subset of events. 
The natural projection from $\Sigma$ to  $\Sigma'$  is a mapping 
$P_{\Sigma'}:\Sigma^* \to \Sigma'^*$ defined recursively by: 
\[
P_{\Sigma'}(\epsilon)=\epsilon 
\text{ and }
P_{\Sigma'}(s\sigma) = 
\left\{ \begin{array}{ll}
P_{\Sigma'}(s)\sigma & \text{if }   \sigma \in  {\Sigma'} \\
P_{\Sigma'}(s) & \text{if }  \sigma \notin  {\Sigma'}  
\end{array} 
\right..
\]
The natural projection is also extended to   $P_{\Sigma'}: 2^{\Sigma^*} \to 2^{\Sigma'^*}$by: for any $L\subseteq\Sigma^*$, $P_{\Sigma'}(L)=\{P_{\Sigma'}(s)\in \Sigma'^*:s\in L\}$.

\subsection{Intruder Model and Current-State Opacity}
In the context of information-flow security analysis, the intruder is usually modeled as a passive observer that (i) knows the model of the system;  and 
(ii) can observe partial behaviors generated by the system. 
Formally, we assume that the event set is partitioned as \vspace{3pt}
\[
\Sigma= \OI\dot{\cup}\UI,  \vspace{3pt}
\]
where $\OI$ and $\UI$ are the sets of observable events and unobservable events of the intruder,  respectively. 
For the sake of simplicity, we use notation $P_a:\Sigma^* \to \OI^*$ to denote the natural projection from $\Sigma$ to $\OI$. Then upon the occurrence of string $s\in \L(G)$, the intruder observes $P_a(s)\in \Sigma_a^*$. 

For any observation $\alpha\in P_a(\L(G))$, the intruder can \emph{estimate} the current-state of the system based on $\alpha$ and the system model $G$. Formally, the \emph{current-state estimate} of the intruder upon the observation of $\alpha$  is defined by \vspace{3pt}
\[
\hat{X}_a(\alpha)= 
\left\{
    \delta(s)\in X:  
 \exists s\in \L(G) \text{ s.t. }   P_a(s)=\alpha 
\right\}. \vspace{3pt}
\] 
 
In the context of standard state-based opacity analysis, it is assumed that the system has a ``secret" modeled by a  set of secret states $X_S\subseteq X$. We denote by $X_{NS}=X\setminus X_S$ the set of non-secret states. Then the system is said to be \emph{current-state opaque} if the intruder can never determine for sure that the system is currently at a secret state. 
In other words, for any string that leads the system to a secret state, there should exist at least one string that leads the system to a non-secret state such that they have the same observation from the intruder's point of view. We review the definition of current-state opacity as follows. 

\begin{definition}[Current-State Opacity]\label{op} 
Given   system $G$, a set of secret states $X_S$ and a set of intruder's observable events  $\OI\subseteq\Sigma$, system $G$ is said to be current-state opaque (w.r.t.\ $\Sigma_a$ and $X_S$) if
\begin{align}
&(\forall s\in \L(G): \delta(s)\in X_S)\\
&(\exists t\in \L(G): \delta(t)\in X_{NS})
[ P_a(s)=P_a(t)]\nonumber
\end{align}
or equivalently, 
\begin{equation}
         (\forall s\in \L(G))
         [\hat{X}_a(P_a(s))\nsubseteq X_S].
\end{equation}
\end{definition}

\section{Notion of High-Order Opacity}\label{sec-notion}

In the standard formulation of current-state opacity, the secret is modeled as the actual behavior of the system. 
In some applications, the user may want to hide its \emph{knowledge about the system}. 
To this end, in this section, we introduce the notion of \emph{high-order opacity}. 
We first present a motivating example and then provide the formal definition. 

\subsection{Motivating Example}

\begin{figure}
  \centering
    \subfigure[Workspace of the robot.\label{motiexample-1}]{\centering
       \definecolor{mygray}{RGB}{192,192,192}
    \definecolor{mycyran}{RGB}{75,211,239}
    \begin{tikzpicture}

    	\draw[fill=mycyran, draw = white] (4.4, 0) -- (8, 0) -- (8, 3.6) -- (4.4, 3.6);
    	
    	\draw[fill=yellow, draw = white] (5.6, 0) -- (6.5, 0) -- (6.5, 0.9) -- (5.6, 0.9);
    	
    	\draw[fill=white, draw = white] (5.6,1.2) -- (6.8,1.2) -- (6.8, 2.1) -- (5.6, 2.1);

    	\draw[fill=yellow, draw = white] (6.5,0) -- (8,0) -- (8, 0.9) -- (6.5, 0.9);
    	\draw[fill=yellow, draw = white] (7.1,0) -- (8,0) -- (8, 1.5) -- (7.1, 1.5);

    	\draw[fill=white, draw = white] (7.1,1.5) -- (8,1.5) -- (8, 3.6) -- (7.1, 3.6);
    	
    	\draw[fill=yellow, draw = white] (4.4, 2.4) -- (6.8, 2.4) -- (6.8, 3.6) -- (4.4, 3.6);
    	
    	\draw[fill=white, draw = white] (4.4,1.8) -- (5.3,1.8) -- (5.3, 2.4) -- (4.4, 2.4);
    	
    	\draw[fill=yellow, draw = white] (4.4,0.9) -- (5.3,0.9) -- (5.3, 1.5) -- (4.4, 1.5);    
    	
    	\draw[fill=white, draw = white] (4.4,0) -- (5.6,0) -- (5.6, 0.9) -- (4.4, 0.9); 
    	
    	\draw[fill=green, draw = white] (5.9,0) -- (6.2,0) -- (6.2, 0.3) -- (5.9, 0.3);
    	\begin{scope}[thick, black, line width=0.35mm, arrows = {-Stealth[scale=0.6]}]
    		\draw (6.05,-0.3) -- (6.05,0);
    	\end{scope}
    	\draw[line width=3pt, black] (6.2, 0.9)--(6.2, 1.2);  
    	\draw[line width=3pt, black] (6.8, 3.0)--(7.1, 3.0);
    	\draw[line width=3pt, black] (5.0, 1.5)--(5.0, 1.8);
    	\draw[fill=red, draw = white] (6.35,0) -- (6.65,0) -- (6.65, 0.3) -- (6.35, 0.3);
    	\draw[fill=red, draw = white] (6.35,0.6) -- (6.65,0.6) -- (6.65, 0.9) -- (6.35, 0.9);
    	\begin{scope}[thick, red, line width=0.35mm, arrows = {-Stealth[scale=0.6]}]
    		\draw (6.35,0.45) -- (6.65,0.45);
    	\end{scope}
    	
    	\draw[fill=red, draw = white] (4.4,2.25) -- (4.7,2.25) -- (4.7, 2.55) -- (4.4, 2.55);
    	\draw[fill=red, draw = white] (5.0,2.25) -- (5.3,2.25) -- (5.3, 2.55) -- (5.0, 2.55);
    	\begin{scope}[thick, red, line width=0.35mm, arrows = {-Stealth[scale=0.6]}]
    		\draw (4.85,2.55) -- (4.85,2.25);
    	\end{scope}

    	\draw[fill=red, draw = white] (4.4,0.75) -- (4.7,0.75) -- (4.7, 1.05) -- (4.4, 1.05);
    	\draw[fill=red, draw = white] (5.0,0.75) -- (5.3,0.75) -- (5.3, 1.05) -- (5.0, 1.05);    	
    	\begin{scope}[thick, red, line width=0.35mm, arrows = {-Stealth[scale=0.6]}]
    		\draw (4.85,0.75) -- (4.85,1.05);
    	\end{scope}
    	
    	\draw[fill=blue, draw = white] (7.1,1.35) -- (7.4,1.35) -- (7.4, 1.65) -- (7.1, 1.65);
    	\draw[fill=blue, draw = white] (7.7,1.35) -- (8.0,1.35) -- (8.0, 1.65) -- (7.7, 1.65);
    	\begin{scope}[thick, blue, line width=0.35mm, arrows = {-Stealth[scale=0.6]}]
    		\draw (7.55,1.35) -- (7.55,1.65);
    	\end{scope}
    	\draw[fill=blue, draw = white] (5.45,0) -- (5.75,0) -- (5.75, 0.3) -- (5.45, 0.3);
    	\draw[fill=blue, draw = white] (5.45,0.6) -- (5.75,0.6) -- (5.75, 0.9) -- (5.45, 0.9);
    	\begin{scope}[thick, blue, line width=0.35mm, arrows = {-Stealth[scale=0.6]}]
    		\draw (5.75,0.45) -- (5.45,0.45);
    	\end{scope}
    	\foreach \x in {4.4,4.7,...,8.3}
    	\draw[densely dotted] (\x, 0)--(\x, 3.6);
    	\foreach \y in {0,0.3,...,3.9}
    	\draw[densely dotted] (4.4, \y)--(8, \y);
    	
    	\node[] [xshift=6.05cm, yshift=0.45cm] (0label) {\footnotesize$0$};
    	\node[] [xshift=4.85cm, yshift=0.45cm] (1label) {\footnotesize$1$};
    	\node[] [xshift=7.55cm, yshift=0.45cm] (2label) {\footnotesize$2$};
    	\node[] [xshift=4.85cm, yshift=1.2cm] (3label) {\footnotesize$3$};
    	\node[] [xshift=7.55cm, yshift=2.55cm] (4label) {\footnotesize$4$};
    	\node[] [xshift=4.85cm, yshift=2.1cm] (5label) {\footnotesize$5$};
    	\node[] [xshift=5.6cm, yshift=3cm] (6label) {\footnotesize$6$};
    	\node[] [xshift=6.2cm, yshift=1.65cm] (7label) {\footnotesize$7$};

    \end{tikzpicture}}
	\subfigure[DFA model of the robot.\label{motiexample-2}]{\centering
   \begin{tikzpicture}[->,>={Latex}, thick, initial text={}, node distance=1.3cm, initial where=below, thick, base node/.style={circle, draw, minimum size=6mm}]  
   \node[state, initial, base node, fill=red] (0) {$0$};
   \node[state, base node, ] (1) [left of=0] {$1$};
   \node[state, base node, fill=red] (2) [right of=0] {$2$};
   \node[state, base node, fill=red] (3) [above of=1] {$3$};
   \node[state, base node, ] (5) [above of=3] {$5$};
   \node[state, base node, fill=red] (6) [right of=5] {$6$};
   \node[state, base node, ] (4) [right of=6] {$4$};
   \node[state, base node, ] (7) [above of=0] {$7$};
   
   \path[<->]
   (0) edge node [xshift=0.2cm] {$g$} (7)
   (3) edge node [xshift=0.2cm] {$g$} (5)
   (4) edge node [yshift=0.2cm] {$g$} (6);
   \path[->]
   (0) edge node [yshift=0.2cm] {$r$} (2)
   (0) edge node [yshift=0.2cm] {$b$} (1)   
   (6) edge node [yshift=0.2cm] {$r$} (5)
   (1) edge node [xshift=0.2cm] {$r$} (3)
   (2) edge node [xshift=0.2cm] {$b$} (4);
   \end{tikzpicture}} \\
  \caption{A motivating example with $\OO=\{r,g\}$ and $\OI=\{b,g\}$.}
	\label{motiexample}
	\vspace{6pt}
\end{figure}
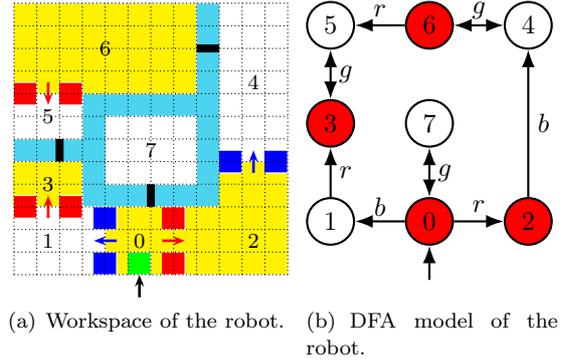

\emph{System Model: }Suppose that there is a robot moving in a workspace
with rivers, bridges and  checkpoints as shown in Figure~\ref{motiexample-1}.  
We assume that the robot can cross the bridges (denoted by black lines) via both directions, but the checkpoints (denoted by red and blue blocks) are only one-way whose directions are specified by arrows in the figure. 
The mobility of the robot can be modeled as DFA $G$ shown in Figure~\ref{motiexample-1}, where states correspond to regions in the workspace and  events $b,r$ and $g$ corresponds to 
``passing a blue checkpoint", ``passing a red checkpoint" and ``crossing the bridge", respectively. 

\emph{User Model: }Suppose that there is a central station that wants to communicate with the robot, e.g, to send commands. 
We assume that  communication signals are only available in those service regions and the central station cannot communicate with the robot  in 
those no-service regions, which are marked by yellow in Figure~\ref{motiexample-1} and are  represented by red states $\{0,2,3,6\}$ in the DFA model.  
Furthermore,  we assume that the central station has sensors placed at each bridge and each red checkpoint, i.e., it can observe the occurrences of events $r$ and $g$. 
In order to make sure that each command sent can be received by the robot \emph{for sure}, the central station will send command to the robot only when it knows for sure that the robot is current at those white regions.

\emph{Intruder Model: }At the same time, we assume that there is an intruder having sensors placed at each bridge and each blue checkpoint, i.e., it can observe the occurrences of events $b$ and $g$.  By knowing the strategy of the central station, the intruder may try to hack to the communication channel between the central station and the robot. Therefore, it will successfully intercept transmitted command when it knows for sure that the central station knows for sure that the robot is at a service region. 
 
\emph{Analysis: }In order to establish communications with the central station,  the robot may choose path $0\overset{g}{\rightarrow}7$, which is the shortest path to reach a service region. Furthermore, the central station will observe event $g$ and upon which it knows for sure that the robot is indeed in a service region. Hence, it will send a command to the robot. However, at the same time, the intruder will also observe event $g$ and upon which it knows for sure that the central station will send a command. Therefore, the command will be intercepted by the intruder, which makes the system not secure. 

To establish a secure communication, the robot can choose path 
$0\overset{r}{\rightarrow}2\overset{b}{\rightarrow}4\overset{g}{\rightarrow}6\overset{r}{\rightarrow}5$ to go to service region $5$. 
Along this path, the central station will observe $rgr$ and it knows for sure that the robot is in a service region. 
However, the intruder will observe $bg$ and it may think that the robot may have chosen path $0\overset{b}{\rightarrow}1\overset{r}{\rightarrow}3\overset{g}{\rightarrow}5$. If it is the case, then the central station will observe $rg$ 
and it cannot distinguish if the robot is at state $5$ or state $6$.  
Therefore, along this path, the central station will know for sure that the robot is in a service region, while the intruder does not know that the central station knows that. This means that the communication in this scenario is secure.  In what follows, we will formulate such a scenario of ``\emph{the intruder does not know that the user knows something}" using the notion of high-order opacity.

\subsection{Knowledge of the User} 
As we discussed in the above motivating example, the user of the system may also have its own observation and based on which it can obtain certain knowledge about the current status of the system.  
To formalize the issue of \emph{knowledge security}, we further assume that the event set is also partitioned as \vspace{3pt}
\[
\Sigma=\OO\dot{\cup}\UO, \vspace{3pt}
\]
where $\OO$ is the set of events that can be observed by the user 
and $\UO$ is the set of events that cannot be observed by the user.   
Similarly, we denote by $P_o:\Sigma^*\to \Sigma_o^*$ the natural projection from $\Sigma$ to $\OO$. 
Also, for any observation of the user $\alpha\in P_o(\L(G))$, we denote by 
$\hat{X}_o(\alpha)$ the current-state estimate of the user upon the observation of $\alpha$. 
Note that,  there is no relationship between the observation of the intruder and the observation of the user, i.e., $\Sigma_a$ and $\Sigma_o$ can be incomparable, because  the intruder and the user may have different sensors monitoring the system.  

For the purpose of decision-making such as control or diagnosis, the user usually wants to determine whether or not it knows some fact based on its partial observation. 
In general, the knowledge of the user can be defined as a predicate on its observations\vspace{3pt}
\[
\texttt{Know}: P_o(\L(G))\to \{\texttt{true},\texttt{false}\}\vspace{3pt}
\]
such that ``$\texttt{Know}(\alpha)=\texttt{true}$" means that the user knows some fact based on observation $\alpha\in P_o(\L(G))$.

In general, by observing $\alpha\in P_o(\L(G))$, the user's complete knowledge about the system is $P_o^{-1}(\alpha)\cap \L(G)$, which is the set of all possible strings consistent with the observation.  In this work, we consider a more concrete type of knowledge of the user called \emph{distinguishability}.   Specifically, we assume that the user is interested in distinguishing certain pairs of states (a.k.a.\ the disambiguation task)\vspace{3pt}
\[
T_{\spec}\subseteq X\times X.  \vspace{3pt}
\]
Specifically, the user wants to distinguish between each pair of states $(x,x')\in T_{\spec}$ and if so, we say that it has sufficient knowledge w.r.t.\ the disambiguation task $T_{\spec}$. 

Then using $T_{\spec}$, we can specify the
 knowledge predicate $\texttt{Know}:\Sigma_o^* \to \{\texttt{true},\texttt{false}\}$ by: 
 for any $\alpha\in P_o(\L(G))$, we have 
$\texttt{Know}(\alpha)=\texttt{true}$ iff\vspace{3pt}
\begin{align}
(\forall s,t \!\in\! \L(G))
[P_o(s)\!=\!P_o(t)\!=\!\alpha\Rightarrow (\delta(s),\delta(t))\!\notin\! T_{\spec}  ]
\end{align}\vspace{5pt}
 or equivalently, 
\begin{equation}
(\hat{X}_o( \alpha )\times \hat{X}_o( \alpha   ))\cap T_{\spec}=\emptyset.\vspace{5pt}
\end{equation}
Intuitively, we have $\texttt{Know}(\alpha)=\texttt{false}$ whenever there are two different strings both have the same observation $\alpha$ but lead to two different states whose pair is in $T_{\spec}$.

\begin{remark}
Although here we consider a specific type of user knowledge captured by disambiguation task $T_{\spec}$, this task is general enough for many practical requirements.   
For example, it can captures ``whether or not the user knows its current location precisely" by setting  $T_{\spec}=\{(x,x')\in X\times X: x\neq x'\}$.  
In general, how to define $T_{\spec}$ is problem dependent and in this work, we will work on the generic knowledge requirement $T_{\spec}$. 
\end{remark}

\subsection{High-Order Opacity}
Before formally introducing the  definition of high-order opacity, we summarize the capabilities of the intruder considered (a.k.a.\ intruder model) as follows
\begin{itemize}
    \item[A1]
    It knows the DFA model  $G$ of the system; \vspace{3pt}
    \item[A2]
    It can observe the occurrence of each event in $\Sigma_a$ generated online; \vspace{3pt}
    \item[A3]
    It knows that the user can observe the occurrence of  each event in $\Sigma_o$ online but itself cannot observe the occurrences of events $\Sigma_o \setminus \Sigma_a$ directly. 
\end{itemize}

The basic idea of high-order opacity still follows the essence of opacity, which is plausible deniability for the secret behavior. 
However, here ``secret" is captured by the knowledge predicate $\texttt{Know}$ rather than secret states $X_S$. 
We require that the intruder should never be able to determine for sure that the user knows something. This leads to the following definition. 

\begin{definition}[High-Order Opacity]\label{highop}
Given  system $G$, a disambiguation task $T_{\spec}\subseteq X\times X$, a set of intruder's observable events  $\OI\subseteq\Sigma$ and a set of user's observable events  $\OO\subseteq\Sigma$, system $G$ is said to be high-order opaque  (w.r.t.\ $T_{\spec}$, $\Sigma_a$ and $\Sigma_o$) if
\begin{align}\label{eq:high}
&(\forall s\in \L(G):\texttt{Know}(P_o(s)))\\
&(\exists t\in \L(G): \neg \texttt{Know}(P_o(t)))
[P_a(s)=P_a(t)].\nonumber
\end{align}
\end{definition} 

The above definition can be understood as follows. 
Let $s\in \L(G)$ be an actual string generated by the system. 
Upon the occurrence of $s$, 
the user observes $P_o(s)$ while the intruder observes $P_a(s)$. 
The user's knowledge is completed determined by $P_o(s)$ but the intruder does not know the user's knowledge perfectly. 
From the intruder's point of view, the user may have observed 
any string in $ P_o(  P_a^{-1}(P_a(s)) \cap \L(G)      ) $, 
and if for any observation $\alpha$ in this language, we have $\texttt{Know}(\alpha)=\texttt{true}$, 
then the intruder knows for sure that predicate $\texttt{Know}$ holds for the user. 
Therefore, the condition in Equation~\eqref{eq:high} can also be expressed equivalently by\vspace{3pt}
\begin{equation}
(\forall \alpha\in P_a(\L(G)))
(\exists \beta\in P_o(  P_a^{-1}( \alpha ) \cap \L(G)  ) )
 \neg\texttt{Know}(\beta).
\end{equation}

We illustrate the notion of high-order opacity by the following example.  

\bexam
\begin{figure}
  \centering
   \begin{tikzpicture}[->,>={Latex}, thick, initial text={}, node distance=1.3cm, initial where=left, thick, base node/.style={circle, draw, minimum size=6mm}]  
   \node[state, initial, base node, ] (0) {$0$};
   \node[state, base node, ] (1) [below of=0] {$1$};
   \node[state, base node, ] (2) [right of=0] {$2$};
   \node[state, base node, ] (3) [below of=2] {$3$};
   \node[state, base node, ] (4) [right of=2] {$4$};
   \node[state, base node, ] (5) [below of=4] {$5$};
   \node[state, base node, ] (6) [right of=4] {$6$};
   \node[state, base node, ] (7) [below of=6] {$7$};
   
   \path[->]
   (0) edge node [yshift=0.2cm] {$c$} (2)
   (2) edge node [yshift=0.2cm] {$b$} (4)
   (4) edge node [yshift=0.2cm] {$d$} (6)
   (0) edge node [xshift=0.2cm] {$a$} (1)
   (1) edge node [yshift=0.2cm] {$b$} (3)
   (3) edge node [yshift=0.2cm] {$b$} (5)
   (5) edge node [yshift=0.2cm] {$a$} (7)
   (7) edge [loop above ] node         [xshift=0.35cm, yshift=-0.1cm]     {$d$} ();
   \draw[->]
   (6) to [bend right=45] node [yshift=0.28cm] {$d$} (4);
   \end{tikzpicture}
   \caption{System $G$ with $\OO=\{b,d\}$ and  $\OI=\{a,b\}$.}\label{example1} 
\end{figure}
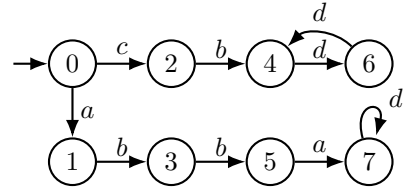

Let us consider system $G$ shown in Figure~\ref{example1}, 
where the user can observe $\Sigma_o=\{b,d\}$ and the intruder can observe $\Sigma_a=\{a,b\}$. 
The disambiguation task of the user is simply to determine the current-state of the system, i.e., 
$T_{\spec}=\{(x,x')\in X\times X: x\neq x'\}$. 
Clearly this system is high-order opaque. 
To see this, we note that the user can determine its current state 
only after it observes the first occurrence of event $d$.  
For string $s=abbad^n$ such that $\texttt{Know}(bbd^n)=\texttt{true}$, the 
intruder observes $P_a(s)=abba$ and it may think that what actually happens is string $s'=abba$
such that $\texttt{Know}(P_o(abba))=\texttt{false}$ because the user cannot distinguish between states $5$ and $7$ by observing $bb$. 
Similarly, for $t=cbd^{2n}$ such that $\texttt{Know}(bd^{2n})=\texttt{true}$,  the 
intruder observes $P_a(t)=b$ and it may think that what actually happens is string $t'=cb$
such that $\texttt{Know}(P_o(cb))=\texttt{false}$ because the user cannot distinguish between states $3$ and $4$ by observing $b$. 
In other word, the intruder can never know that the observer knows the current state of the system $G$.
\eexam

\bremark
It is worth noting that  high-order opacity only requires that \emph{whenever} the user has the knowledge to distinguish $T_{\spec}$, the intruder does not know this fact. This definition itself does not require that the user can always or will eventually have this knowledge. This is essentially a utility requirement, which is a different consideration apart from the security consideration.  In other words, it is even possible that the intruder knows the secret information (distinguishing $T_{\spec}$), but it does not know whether or not the user also knows this secret information. 
For instance, in Figure~\ref{example1}, if the intruder observes $abba\in P_a(\L(G))$, it can determine for sure that the system  is currently at state $7$. However, it can not determine whether the user knows the current state because $\hat{X}_o(P_o(abba))=\{5,7\}$, $\hat{X}_o(P_o(abbad))=\{7\}$ and $P_a(abba)=P_a(abbad)=abba$.
This scenario does not violate the requirement of high-order opacity according to the definition. 
\eremark

\begin{remark}\label{remark:compare}
The new notion of high-order opacity subsumes the standard current-state opacity as defined in Definition~\ref{op}.  
To see this, we can set $\Sigma_o=\Sigma$ and $T_{\spec}=\{ (x,x)\in X: x\in X_{NS} \}$ and then high-order opacity becomes current-state opacity.  
This is because $T_{\spec}$ can never be distinguished when the system is not at a secret state. Furthermore, since the user knows the current state perfectly, $\texttt{Know}(\alpha)=\texttt{true}$ is equivalent to the fact that the system is currently at a secret state.  
This reduction implies that high-order opacity is more general than current-state opacity. 
Since it is known that the verification of current-state opacity is PSPACE-complete \citep{balun2020opacity}, we can conclude immediately that verifying high-order opacity is at least PSPACE-hard. 
\end{remark}

\section{Verification of High-order Opacity}\label{sec-verify}

In this section, we present two algorithms for the verification of the proposed notion of high-order opacity; one is based on the construction of double-observer and the other is based on the construction of state-pair-observer. 

\subsection{Double-Observer Approach} 
Let $G=(X,\Sigma,\delta,x_0)$ be a DFA and $\Sigma'\subseteq \Sigma$ be a subset of observable event. 
Then the observer automaton of $G$ w.r.t.\ $\Sigma'$ is a new DFA\vspace{3pt}
\[
\Obs_{\Sigma'}(G)=(Q, \Sigma, f, q_0), \vspace{3pt}
\]
where  
$Q\subseteq 2^X \setminus \emptyset$ is the set of states,  
$q_0=\{ \delta(w)\in X:   w\in  (\Sigma\setminus \Sigma')^*   \}$ is the   initial state, 
and 
$f:Q\times \Sigma\to Q$ is the deterministic transition function defined by: 
for any $q\in Q$ and $\sigma\in\Sigma$,  we have
\begin{itemize}
    \item 
    if $\sigma \in \Sigma'$, then  \vspace{3pt}
    \[
f(q,\sigma)=
\{
\delta(x,\sigma w) \in X:   x\in q, w \in (\Sigma\setminus \Sigma')^*   
\} \vspace{3pt}
\]
\item 
if $ \sigma\in \Sigma\setminus \Sigma'$ and there exists $x\in q$ such that $\delta(x,\sigma)!$, then 
$f(q,\sigma)=q$. 
\end{itemize}
Intuitively, the observer automaton  tracks all possible current states of system  based on observations in  $\Sigma'^*$. Specifically, for any $\alpha\in P_{\Sigma'}(\L(G))$, $f(q_0,\alpha)$ is the current-state estimate of $\alpha$ w.r.t.\ $\Sigma'$. For technical purposes, here we further add self-loops at each state for those feasible but unobservable events in $\Sigma\setminus \Sigma'$, which ensures that $\L(G)\subseteq\L(\Obs_{\Sigma'}(G))$. 

Note that the knowledge of the user is described by distinguishability which is based on the current-state estimation.  Therefore, we can build the observer automaton w.r.t.\ $\Sigma_o$ denoted by  $\Obs_{\Sigma_o}(G)=(Q_o,\Sigma,f_o,q_{o,0})$ to capture this issue. Specifically, since for any $\alpha\in P_o(\L(G))$,  the knowledge predicate $\texttt{Know}$ holds iff $(f_o(\alpha)\times f_o(\alpha))\cap T_\spec=\emptyset$. 
Therefore, we define the set of observer states for which the knowledge predicate holds as
\[
Q_{o,S}=\{  q\in Q_o: (q\times q)\cap T_\spec= \emptyset\}.
\]

For high-order opacity, we need to further think from the intruder's point of view, i.e., how the intruder estimates the state estimate of the user. Therefore, we need to further build the observer automaton of $\Obs_{\Sigma_o}(G)$ w.r.t.\ $\Sigma_a$, which is referred to as the \emph{double-observer}, as follows 
\[
\Obs_D(G):=\Obs_{\Sigma_a}(\Obs_{\Sigma_o}(G))=(Q_D,\Sigma,f_D,q_{D,0})
\]

Intuitively, the double-observer $\Obs_D(G)$  tracks all the possible observer states in  $\Obs_o(G)$ based on another event set   $\OI$. Also we can get that for any $s,t\in\L(G)$, we have $f_D(s)=f_D(t)$ if $P_a(s)=P_a(t)$ from the definition of the transition function of $\Obs_D(G)$. In other word, the states of $\Obs_D(G)$ is actually the current state estimate of observer $\Obs_o(G)$ from the intruder's point of view. Using $\Obs_D(G)$, we can easily check high-order opacity by the following theorem.

\btheorem\label{theo1}
System $G$ is high-order opaque  (w.r.t.\ $T_{\spec}$, $\Sigma_a$ and $\Sigma_o$) iff
$\forall q\in Q_D: q\nsubseteq Q_{o,S}$.  
\etheorem

\begin{pf}
($\Rightarrow$)
By contraposition.  
Suppose that there exists a state $q\in Q_D$ such that $q\subseteq Q_{o,S}$. 
Then for any $s\in\L(G)$ such that  $f_D(q_{D,0},s)=q$ we have $f_o(q_{o,0},s)\in Q_{o,S}$, which means that  $\texttt{Know}(P_o(s))=\texttt{true}$. Also,  for any other  string  $t\in\L(\Obs_D(G))$ such that $P_a(s)=P_a(t)$, we have $f_D(q_{D,0},s)=f_D(q_{D,0},t)$. Since $q\subseteq Q_{o,S}$, we know  that there exists no string $t\in\L(G)$ such that $f_o(q_{o,0},t)\notin Q_{o,S}$ and $P_a(t)=P_a(s)$. In other words, for any $t\in \L(G)$ such that $P_a(s)= P_a(t)$, we have  $\texttt{Know}(P_o(t))=\texttt{true}$, which means the system is not high-order opaque.

($\Leftarrow$) 
Also by contraposition. 
Suppose that the system is not high-order opaque, which means that there exists a string 
$s\in\L(G)$ such that $\texttt{Know}(P_o(s))=\texttt{true}$. 
Therefore, we have $f_o(q_{o,0},s)\in Q_{o,S}$ and there exists no string $t\in\L(G)$ such that  $\texttt{Know}(P_o(t))=\texttt{false}$ and $P_a(t)=P_a(s)$. This means that for any $t\in\L(G)$ satisfies $P_a(t)=P_a(s)$, we have $f_o(q_{o,0},t)\in Q_{o,S}$. Therefore, we have $f_D(q_{D,0},s)\subseteq Q_{o,S}$.
\hfill $\qed$
\end{pf}

The above theorem immediately suggests Algorithm~1 for verifying high-order opacity. 
Specifically, we need to build the double-observer and check whether or not it contains a state such that the knowledge predicate holds for  each element in it.
In the worst-case, the doubly-observer contains $2^{2^{|X|}}$ states and $|\Sigma|2^{2^{|X|}}$ transitions. 
Therefore, the overall complexity of Algorithm~1 is doubly exponential in the size of $G$.  

\begin{algorithm}
\caption{High-Order-Opa-Dou-Obs} 
\hspace*{0.02in} {\bf Input:} 
$G$, $T_\spec$, $\Sigma_a$, $\Sigma_o$\\
\hspace*{0.02in} {\bf Output:} 
High-order opaque   or not
\begin{algorithmic}[1]
\State Build $\Obs_o(G)=(Q_o,\Sigma,f_o,q_{o,0})$
\State Build $\Obs_D(G)=(Q_D,\Sigma,f_D,q_{D,0})$

\ForAll{$p\in Q_D$}
    \If{$p\subseteq Q_{o,S}$}
        \State \Return $G$ is not high-order opaque
    \EndIf
\EndFor
\State \Return $G$ is high-order opaque
\end{algorithmic}
\end{algorithm}

Although the worst-case complexity of Algorithm~1 is doubly-exponential, empirical studies show that, in many systems, the exponential state-space explosion in the observer construction does not really occur \citep{clavijo2017empirical}. This is why we still choose to present this algorithm. We illustrate Algorithm 1 using the following example.

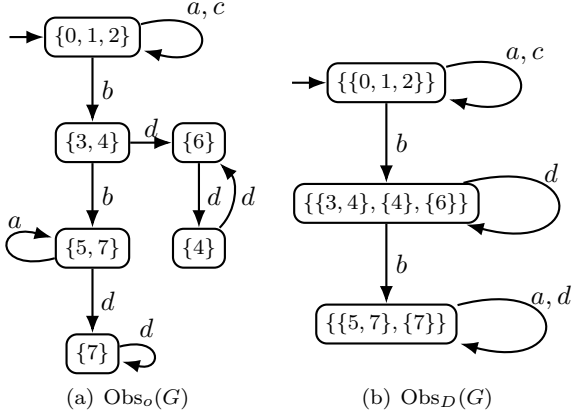
\begin{figure}
  \centering
    \subfigure[$\Obs_o(G)$\label{fig2-1}]{\centering
   \begin{tikzpicture}[->,>={Latex}, thick, initial text={}, node distance=1.4cm, initial where=left, thick, base node/.style={rectangle, rounded corners, draw, minimum size=4mm, font=\footnotesize}]  
   \node[initial, state, base node, rectangle, rounded corners, ] (0) {$\{ 0,1,2 \}$} ;
   \node[state, base node, rectangle, rounded corners, ] [below of=0] (1) {$\{3,4\}$} ;
   \node[state, base node, rectangle, rounded corners, ] [right of=1] (2) {$\{6\}$} ;
   \node[state, base node, rectangle, rounded corners, ] [below of=1] (3) {$\{5,7\}$} ;
   \node[state, base node, rectangle, rounded corners, ] [right of=3] (4) {$\{4\}$} ;
   \node[state, base node, rectangle, rounded corners, ] [below of=3] (5) {$\{7\}$} ;
   
   \path[->]
   (0) edge node [xshift=0.2cm] {$b$} (1)
   (1) edge node [yshift=0.2cm] {$d$} (2)
   (1) edge node [xshift=0.2cm] {$b$} (3)
   (2) edge node [xshift=0.2cm] {$d$} (4)
   (3) edge node [xshift=0.2cm] {$d$} (5)
   (0) edge [loop right] node         [xshift=-0.3cm, yshift=0.35cm]     {$a,c$} ()
   (3) edge [loop left] node         [xshift=0.3cm, yshift=0.35cm]     {$a$} ()
   (5) edge [loop right] node         [xshift=-0.3cm, yshift=0.35cm]     {$d$} ();
   \draw[->]
   (4) to [bend right=45] node [xshift=0.2cm] {$d$} (2);
   \end{tikzpicture}}
	\subfigure[$\Obs_D(G)$	\label{fig2-2}]{\centering
   \begin{tikzpicture}[->,>={Latex}, thick, initial text={}, node distance=1.6cm, initial where=left, thick, base node/.style={rectangle, rounded corners, draw, minimum size=4mm, font=\footnotesize}]  
   \node[initial, state, base node, rectangle, rounded corners, ] (0) {$\{\{ 0,1,2 \}\}$} ;
   \node[state, base node, rectangle, rounded corners, ] [below of=0] (1) {$\{\{3,4\},\{4\},\{6\}\}$} ;
   \node[state, base node, rectangle, rounded corners, ] [below of=1] (2) {$\{\{5,7\},\{7\}\}$} ;
   \path[->]
   (0) edge node [xshift=0.2cm] {$b$} (1)
   (1) edge node [xshift=0.2cm] {$b$} (2)
   (0) edge [loop right] node         [xshift=-0.3cm, yshift=0.35cm]     {$a,c$} ()
   (1) edge [loop right] node         [xshift=-0.3cm, yshift=0.4cm]     {$d$} ()
   (2) edge [loop right] node         [xshift=-0.3cm, yshift=0.35cm]     {$a,d$} ();
   \end{tikzpicture}} \\
  \caption{Two observer automaton for $G$.	}
	\label{fig2}
	\vspace{6pt}
\end{figure}

\bexam
Again, let us consider system $G$  presented in Example 1. 
The observer $\Obs_o(G)$ w.r.t.\ event set $\Sigma_o$ is shown in Figure~\ref{fig2-1}. 
Since we consider knowledge task $T_{\spec}=\{(x,x')\in X\times X: x\neq x'\}$, 
for any string $s\in \L(G)$,  we have $\texttt{Know}(P_o(s))=\texttt{true}$ 
iff $f_o(q_{o,0},s)$ is a singleton. 
Therefore, we have  $Q_{o,S}=\{\{4\},\{6\},\{7\}\}$. 
Based on  observer $\Obs_o(G)$, we further build the double-observer $\Obs_D(G)$ w.r.t.\ $\Sigma_a$ as shown in Figure~\ref{fig2-2}. 
For each state $q\in Q_D$ in it, we see that $q$ always contains an element not in $Q_{o,S}$. 
Therefore, by Theorem~\ref{theo1}, we conclude that $G$ is high-order opaque. 
This conclusion is consistent with our analysis in Example 1.

\eexam

\subsection{State-Pair-Observer Approach}
The double-observer approach uses the subset construction technique to capture information uncertainties for both the user and the intruder. However, since we use the subset construction twice, the double-observer is doubly-exponential in the size of the plant. 
Here, we note that, from the user's point of view, it is only interested in whether or not it can distinguish all pairs in $T_{\spec}$. Therefore, from the intruder's point of view, it suffices to estimate the set of all state pairs the user cannot distinguish. This leads to the  \emph{state-pair-observer} defined as follows.  

\begin{definition}[State-Pair-Observer]
Given  DFA $G=(X,$ $\Sigma,\delta,x_0)$, observable events of the user $\Sigma_o\subseteq \Sigma$ and   observable events of the intruder $\Sigma_a\subseteq \Sigma$, the state-pair-observer is a new DFA
\[
\Obs_V(G)=( Q_V, \Sigma_a,  f_V, q_{0,V}), 
\] 
where
\begin{itemize}
    \item 
    $Q_V \subseteq 2^{X\times X}\setminus\emptyset$ is the set of states;
    \item 
    $\Sigma_a$ is the set of events;  
    \item 
    $f_V: Q\times \Sigma_a \to Q$ is the deterministic transition function defined by: 
    for any $q\in Q_V, \sigma\in \Sigma_a$, we have
    \begin{align}
     &f_V(q,\sigma)=     \nonumber\\
    & \left\{
     (x_1',x_2') :
     \begin{array}{cc}
       \exists (x_1,x_2)\in q, \exists w\in \Omega_{ua}(q,\sigma), \\ \exists w_1',w_2'\in  \Sigma^* \text{ s.t.} \\
       P_o(\sigma w)=P_o(w_1')=P_o(w_2')\text{ and} \\
       (x_1',x_2')=(\delta(x_1,w_1'),\delta(x_2,w_2'))
    \end{array} 
     \right\}, 
    \nonumber
    \end{align}
    where   $\Omega_{ua}(q,\sigma)$ is the set of feasible unobservable strings defined by:
    \[
    \Omega_{ua}(q,\sigma)= \{ w\!\in \!\Sigma_{ua}^* : \exists (x,x)\!\in\! q \text{ s.t. } \delta(x,\sigma w)!\}.
    \]
    \item 
    $q_{0,V}$ is the unique initial state defined by:
     \begin{align}
     &q_{0,V}=     \nonumber\\
    & \left\{\!\!
     (x_1',x_2') :\!\!
     \begin{array}{cc}
       \exists w\! \in\!  \Sigma_{ua}^* \cap \L(G),  
       w_1',w_2'\! \in\!   \L(G) \text{ s.t. } \\
       P_o(w)=P_o(w_1')=P_o(w_2')\text{ and} \\
       (x_1',x_2')=(\delta(w_1'),\delta(w_2'))
    \end{array} \! \! 
     \right\}
    \nonumber
    \end{align}
\end{itemize}
\end{definition}

Intuitively, the state-pair-observer estimates from the intruder's point of view based on $\Sigma_a$. 
However, instead of estimating the state-estimate of the user as the case of double-observer, it estimates the set of state pairs that can not be distinguished by the user. 
Specifically, suppose that the current estimate of the intruder is $q\in 2^{X\times X}$, which means that it thinks that the user cannot distinguish state pair $(x_1,x_2)\in q$. 
Then by observing a new event $\sigma \in \Sigma_a$, the intruder needs to update its estimate by considering all strings with unobservable tail $w\in \Omega_{ua}(q,\sigma)$, where state pair $(x,x)\in q$ represents a possible actual state $x$ of the system. 
Note that, for each actual string $\sigma w$ in the system, the user has its own observation according to $\Sigma_o$. 
Then for strings $w_1\in \L(G,x_1)$ and $w_2\in \L(G,x_2)$, if $P_o(w_1)=P_o(w_2)=P_o(\sigma w)$, 
then the user again cannot distinguish between state $\delta(x_1,w_1)$ and state $\delta(x_2,w_2)$.

The following result says that the proposed state-pair-observer indeed captures all pairs of states the user cannot distinguish based on the observation of the intruder. 

\bproposition\label{proposition1}
For any $s\in \L(G)$, the state reached by $P_a(s)$ in $\Obs_V(G)$ satisfies the following:
\begin{align}
   & f_V(q_{0,V},P_a(s))=    \nonumber  \\
   & \left\{
   (\delta(w_1'),\delta(w_2'))  :
    \begin{array}{cc}
       \exists  t, w_1', w_2'\in  \L(G), \\
       \text{s.t. } P_a(s)=P_a(t)\text{ and}\\ 
           P_o(t)=P_o(w_1')=P_o(w_2') 
     \end{array} 
         \right\}
    \nonumber
\end{align}
\eproposition

\begin{pf}
For any $s\in\L(G)$, 
we can write in the form of $s=w_0 \sigma_1 w_1\cdots\sigma_n w_n$, where 
$\sigma_i\in \Sigma_a$ and $w_i \in \Sigma_{ua}^*$. 
First, we show by induction that for any $s\in \L(G)$, $P_a(s)$ is always well-defined in $\Obs_V(G)$.  
To this end, we show by induction that for any string $s\in \L(G)$, $(\delta(s),\delta(s))\in f_V(q_{0,V},P_a(s))$. For $n=0$, i.e., $P_a(s)=\epsilon$, since $(x_0,x_0)\!\in\! q_{0,V}$  and $\delta(w_0)!$, clearly we have $(\delta(s),\delta(s))\in  q_{0,V}$. 
We assume that this holds for $n=i$ and  for $n=i+1$, i.e., 
$P_a(s)=\sigma_1 \cdots \sigma_{i+1}$. 
For the sake of simplicity, we write the corresponding state as $x_{i+1}\!=\!\delta(x_i, \sigma_{i+1} w_{i+1})\!=\! \delta(w_0\sigma_1 w_1 \cdots \sigma_{i+1} w_{i+1})$. 
According to the definition of $f_V(q_{i,V},\sigma_{i+1})$, where $q_{i,V}=f_V(q_{0,V}, \sigma_1...\sigma_i)$,  since $w_{i+1}\in \Omega_{ua}(q_{i,V},\sigma_{i+1})$,   we have $(x_{i+1},x_{i+1})\!\in\! f_V(q_{i,V},\sigma_{i+1}) \!=\! f_V(q_{0,V}, \sigma_1 \cdots \sigma_{i+1})$.
This implies that $P_a(s)$ is always well-defined in $\Obs_V(G)$ for any $s\in \L(G)$. 

Now, we  prove Proposition~\ref{proposition1}  also by induction for on $n=|P_a(s)|$.

\emph{Induction Basis: } 
For $n=0$, i.e., $P_a(s)=\epsilon$, Proposition~\ref{proposition1} is actually the reformulation of definition of $q_{0,V}$.  Therefore, the induction basis holds. 

\emph{Induction Step: } 
Now we assume that for $n=i$. i.e., $P_a(s)=\sigma_1 \cdots \sigma_i$,  Proposition~\ref{proposition1} holds.
Then for $n=i+1$, according to the definition of $f_V(q,\sigma_{i+1})$, we have
    \begin{align}
     &f_V(f_V(q_{0,V},\sigma_1\cdots\sigma_i),\sigma_{i+1})=     \nonumber\\
    & \left\{\!\!
     (x_1',x_2') :\!\!\!\!\!
     \begin{array}{cc}
       \exists (x_{1i},x_{2i})\!\in\!f_V(q_{0,V},\sigma_1...\sigma_i), \\ \exists w\!\in\! \Omega_{ua}(f_V(q_{0,V},\sigma_1...\sigma_i),\sigma_{i+1}), \\ \exists w_{1(i+1)}',w_{2(i+1)}'\!\in\!  \Sigma^* \text{ s.t.} \\
       P_o(\sigma_{i+1} w)\!=\!P_o(w_{1(i+1)}')\!=\!P_o(w_{2(i+1)}')\text{ and} \\
       (x_1',x_2')=(\delta(x_{1i},w_{1(i+1)}'),\delta(x_{2i},w_{2(i+1)}'))
    \end{array} \!\!\!
     \right\}
    \nonumber
    \end{align}    
Let $w_1'=w_{10}'w_{11}'\cdots w_{1(i+1)}'$ and $w_2'=w_{20}'w_{21}'\cdots w_{2(i+1)}'$. Clearly, we have   $w_1',w_2'\in \L(G)$ since $w_{1(j+1)}'\in \L(G,x_{1j})$, $w_{2(j+1)}'\in \L(G,x_{2j})$ for all $j\leq i, j\in\mathbb{N}$, and $w_{10}', w_{20}'\in \L(G)$. By using the induction hypothesis,  we further have 
    \begin{align}
        & f_V(q_{0,V},\sigma_1\cdots\sigma_{i+1})=    \nonumber  \\
        & \left\{
         (\delta(w_1'),\delta(w_2')) :
         \begin{array}{cc}
           \exists t_{i+1}, w_1',w_2'\in  \L(G) \text{ s.t.} \\ P_a(s)=P_a(t_{i+1}) \text{ and}\\ 
           P_o(t_{i+1})=P_o(w_1')=P_o(w_2')
        \end{array} 
         \right\}
    \nonumber
    \end{align}
This completes the induction step. 
\hfill $\qed$
\end{pf}

Then for any observer state $q\in 2^{X\times X}$, we say $q$ is a \emph{secret revealing state} if any state pair $(x_1,x_2)\in q$ in it is not in $T_{\spec}$, and we define \vspace{3pt}
\[
Q_{V,S}= \{ q\in Q_V:    q\cap T_{\spec} =\emptyset \}\vspace{3pt}
\]
as the set of secret revealing states. Recall that each pair $(x_1,x_2)\in q$ is a pair of states the user \emph{cannot} distinguish. 
Therefore, a secret revealing state means that the intruder knows for sure that the user \emph{can} distinguish task $T_{\spec}$; 
hence, the knowledge secret is revealed. This leads to the following main theorem.  

\btheorem\label{theo2}
System $G$ is high-order opaque (w.r.t.\ $T_{\spec}$, $\Sigma_a$ and $\Sigma_o$)  iff 
$Q_{V,S}=\emptyset$. 
\etheorem

\begin{pf}
($\Rightarrow$)
By  contraposition. 
Suppose that $Q_{V,S}\ne\emptyset$, i.e., there exists a state $q\in Q_V$ such that $q\cap T_\spec=\emptyset$.
Let us consider string $s\in\L(G)$ such that  $f_V(q_{0,V},P_a(s))=q$. According to Proposition~\ref{proposition1},  for any $t, w_1, w_2\in \L(G)$ such that $P_o(t)=P_o(w_1)=P_o(w_2)$ and $P_a(s)=P_a(t)$, we have $(\delta(w_1),\delta(w_2))\notin T_\spec$. Furthermore, the product of the state estimation of $P_o(s)$ is
    \begin{align}
     &\hat{X}_o(P_o(s))\times \hat{X}_o(P_o(s))=     \nonumber\\
    & \left\{ 
     (\delta(w_1),\delta(w_2)) : 
     \begin{array}{cc}
       \exists w_1,w_2\! \in\!   \L(G) \text{ s.t.} \\
       P_o(s)=P_o(w_1)=P_o(w_2)
    \end{array}  
     \right\}.
    \nonumber
    \end{align}
Therefore, $\forall t\in \L(G):P_a(t)=P_a(s)$, $\texttt{Know}(P_o(t))=\texttt{true}$. This means $G$ is not high-order opaque.

($\Leftarrow$) 
Also by contraposition.  
Suppose that the system is not high-order opaque. 
Then there exists a string $s\in\L(G)$ such that 
\[ 
\forall t\in\L(G): P_a(t)=P_a(s)\Rightarrow  \texttt{Know}(P_o(t))=\texttt{true}
\]
That is,  
\begin{align}
& (\forall t\in\L(G): P_a(t)=P_a(s)) \nonumber\\\
&(\forall  w_1,w_2  \in\L(G): P_o(t)=P_o(w_1)=P_o(w_2)) \nonumber\\\
&[  (\delta(w_1),\delta(w_2))\notin T_\spec 
]\nonumber
\end{align}
Then by considering string $P_a(s)\in \L(\Obs_V(G))$, 
according to Proposition~\ref{proposition1}, 
we  know that    $\forall (x_1,x_2)\in f_V(q_{0,V},P_a(s)):(x_1,x_2)\notin T_\spec $. 
This means that $f_V(q_{0,V},$ $P_a(s))\in Q_{V,S}$, i.e., $Q_{V,S}\ne\emptyset$.
\hfill $\qed$
\end{pf}

The above theorem suggests immediately how to use state-pair-observer for verifying high-order opacity. 
The procedure is summarized in Algorithm 2, which is essentially a reachability search in $\Obs_V(G)$. 
Note that there are at most $2^{|X|^2}$ states and $|\Sigma_a|2^{|X|^2}$ transitions in the state-pair-observer. 
Therefore, the worst-case complexity of Algorithm~2 is  $O(|\Sigma_a|2^{|X|^2})$, which is single-exponential in the size of $G$. 
As we have discussed in Remark~\ref{remark:compare}, verifying high-order complexity is at least PSPACE-hard since it subsumes the standard notion of current-state opacity. Then with this single-exponential upper-bound, we can further conclude that verifying high-order complexity is at actually PSPACE-complete.

\begin{algorithm}
\caption{High-Order-Opa-Sta-Pai-Obs} 
\hspace*{0.02in} {\bf Input:} 
$G$, $T_\spec$, $\Sigma_a$, $\Sigma_o$\\
\hspace*{0.02in} {\bf Output:} 
High-order opaque   or not
\begin{algorithmic}[1]
\State Build $\Obs_V(G)=(Q_V, \Sigma_a, f_V, q_{0,V})$
\ForAll{$q\in Q_V$}

    \If{$q\in Q_{V,S}$}
        \State \Return $G$ is not high-order opaque
    \EndIf
\EndFor
\State \Return $G$ is high-order opaque
\end{algorithmic}
\end{algorithm}

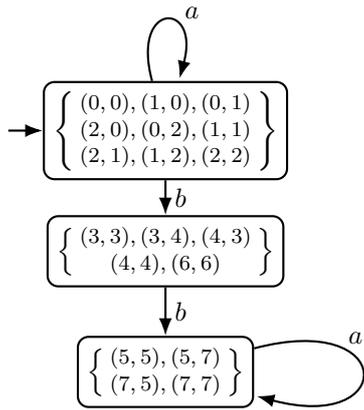
\begin{figure}
  \centering
	\begin{tikzpicture}[->,>={Latex}, thick, initial text={}, node distance=1.6cm, initial where=left, thick, base node/.style={rectangle, rounded corners, draw, minimum size=4mm, font=\footnotesize}]  
   \node[initial, state, base node, rectangle, rounded corners, ] (0) {
   $ \left\{ \begin{array}{c c c} (0,0), (1,0), (0,1)  \\ (2,0), (0,2), (1,1)  \\ (2,1), (1,2), (2,2) \end{array} \right\} $ } ;
   \node[state, base node, rectangle, rounded corners, ] [below of=0] (1) { $ \left\{ \begin{array}{c c} (3,3), (3,4), (4,3)  \\ (4,4), (6,6) \end{array} \right\} $ } ;
   \node[state, base node, rectangle, rounded corners, ] [below of=1] (2) { $ \left\{ \begin{array}{c c} (5,5), (5,7)  \\ (7,5), (7,7) \end{array} \right\} $ } ;
   \path[->]
   (0) edge node [xshift=0.2cm] {$b$} (1)
   (1) edge node [xshift=0.2cm] {$b$} (2)
   (0) edge [loop above] node         [xshift=0.35cm, yshift=-0.1cm]     {$a$} ()
   (2) edge [loop right] node         [xshift=-0.3cm, yshift=0.45cm]     {$a$} ();
   \end{tikzpicture}
  \caption{State-pair-observer $Obs_V(G)$ for $G$.}
	\label{fig:sp-obs}
	\vspace{6pt}
\end{figure}

\bexam
Let us still consider system $G$ shown in Figure~\ref{example1} with $T_{\spec}=\{(x,x')\in X\times X: x\neq x'\}$. 
This state-pair-observer is shown in Figure~\ref{fig:sp-obs}. 
Initially, we have $P_a(\epsilon)=P_a(c)=\epsilon$. 
Note that we also have $P_o(\epsilon)=P_o(c)=\epsilon$
and $P_o^{-1}(\epsilon )\cap \L(G)=\{\epsilon, a,c\}$, 
i.e., the intruder thinks that user cannot distinguish 
states $\delta(\epsilon)=0,\delta(a)=1$ and $\delta(c)=2$. 
Therefore, the $3\times 3$ combinations of the state pairs gives the initial-state in $\Obs_V(G)$. 
Also, for example, consider string $s\!=\!cbdd$, where  $P_a(s)\!=\!b$. 
We have $P_a^{-1}(P_a(s))\cap \L(G)=\{  cbd^n  \}$. 
For strings  $t_1\!=\!cbd^n$ where $n\geq 1$, we have $P_o^{-1}(P_o(cbd^n))\cap \L(G)= \{cbd^n\}$, i.e., the user can perfectly determine the state, and  corresponding state pairs the user cannot distinguish are $(4,4)$ and $(6,6)$. 
For strings $t_2\!=\!cb$, we have $P_o^{-1}(P_o(cb))\cap \L(G)= \{ab, cb\}$. Therefore, the state pairs the user cannot distinguish are $(3,3),(4,4),(3,4)$ and $(4,3)$.
This is why we have $f_V(q_{0,V},P_a(s))\!=\!\{(3,3),(4,4),(3,4),(4,3),(6,6)\}$.
Note that, for each state $q\in Q_V$, $q$ contains a state pair $(x_1,x_2)$ such that $x_1\neq x_2$, which means that 
$q\cap T_\spec\neq \emptyset$. Intuitively, the means that the intruder thinks that the user may not be able to distinguish between different states $x_1$ and $x_2$. 
Since   $Q_{V,S}=\emptyset$, according to Theorem~\ref{theo2}, we conclude that the system $G$ is high-order opaque, which is also consistent with our previous analysis. 

\eexam

\section{Conclusion}\label{sec-con}
In this paper, we investigated information-flow security properties of partially-observed DES from a new angle by considering secret as the user's  knowledge of some fact of importance.  
We proposed the new notion of high-order opacity to capture the knowledge security requirement, i.e.,  the intruder can never know that the user knows something for sure. We showed that the new notion of high-order opacity subsumes the standard notion of current-state opacity. Effective algorithms were also provided for verifying this new notion. 

This paper makes the first step towards the framework of knowledge security and there are several on-going and future directions. First, in this paper, we essentially consider the \emph{current knowledge} of the user. In some applications, however, what the user wants to hide may be the fact that he \emph{knew something at some previous instant}. This is analogous to the difference between current-state opacity and infinite/$K$-step opacity. Also, this paper only considers the verification of high-order opacity. When the system is verified to be non-opaque, one may further want to enforce high-order opacity using supervisory control \citep{dubreil2010supervisory} or insertion functions \citep{ji2019enforcing}. These problems are still under investigation. 
\bibliography{ifacconf}

\end{document}